# Earthquake Counting Method
# for Spatially Localized Probabilities:
# Challenges in Real-Time Information Delivery

*by*

James R Holliday
Department of Physics
University of California, Davis, CA

William R Graves
Open Hazards Group
Davis, CA

John B Rundle
Departments of Physics and Geology
University of California, Davis, CA
*and*
The Santa Fe Institute
Santa Fe, NM

Donald L Turcotte
Department of Geology
University of California, Davis, CA

# Abstract

We develop and implement a new type of global earthquake forecast. Our forecast is a perturbation on a smoothed seismicity (Relative Intensity) spatial forecast combined with a temporal time-averaged ("Poisson") forecast. A variety of statistical and fault-system models have been discussed for use in computing forecast probabilities. An example is the Working Group on California Earthquake Probabilities, which has been using fault-based models to compute conditional probabilities in California since 1988. This WGCEP model is based on defining model earthquake faults, and assigning probabilities to various rupture scenarios. The results of these calculations are used as inputs to damage and loss models used to compute earthquake insurance rates. Another example of a forecast is the Epidemic-Type Aftershock Sequence (ETAS), which is based on the Gutenberg-Richter (GR) magnitude-frequency law, the Omori aftershock law, and Poisson statistics. In a more recent paper, we have taken a different approach. The idea is based on the observation that GR statistics characterize seismicity for all space and time. Small magnitude event counts (quake counts) are used as "markers" for the approach of large events. More specifically, if the GR b-value = 1, then for every 1000 M>3 earthquakes, one expects 1 M>6 earthquake. So if ~1000 M>3 events have occurred in a spatial region since the last M>6 earthquake, another M>6 earthquake should be expected soon. In physics, event count models have been called natural time models, since counts of small events represent a physical or natural time scale characterizing the system dynamics. In a previous paper, we used conditional Weibull statistics to convert event counts into a temporal probability for a given fixed region. In the present paper, we dispense with a fixed region, and develop a method to compute these Natural Time Weibull (NTW) forecasts on a global scale, using an internally consistent method, in regions of arbitrary shape and size. We develop and implement these methods on a modern web-service computing platform, which can be found at www.openhazards.com and www.quakesim.org. We also discuss constraints on the User Interface (UI) that follow from practical considerations of site usability. Among the results we find that the Japan region is at serious risk for a major (M>8) earthquake over the next year or two, a result that also follows from considering completeness of the Gutenberg-Richter relation.

# 1. Introduction

**1.1 Background.** The Working Group on California Earthquake Probabilities [1] computes the official earthquake forecast for the state of California. This is a (primarily) fault-based forecast that relies on crustal deformation, paleo-seismology, seismicity, and related data, to produce a model for use in earthquake insurance and formulation of building codes. In the past, the methods of computation have used several types of statistical distributions, including Poisson, log-normal, and Brownian Passage Time (Evans et al., 1993; Matthews et al., 2002) to convert fault-recurrence times into earthquake probabilities.

Expert elicitation is used as a method to construct and weight logic trees for the forecast selection process ([1]; Field et al., 2009). The first working group report was published in 1988, and the most recent report was published in 2009 (Field et al., 2009). The fault-based forecast is particularly useful for models of structural damage and loss developed by the commercial companies EQECAT, RMS, and AIR, as well as the HAZUS model available from FEMA. A characteristic common to all the WGCEP models is that a forecast of this type requires several years of effort by large numbers of scientists and engineers to produce. Future forecast methods may adopt alternative approaches as well, with the development and use of topologically realistic numerical earthquake simulations (e.g., Rundle et al., 2005; also ref. [1]).

A group using a somewhat similar approach is the Global Earthquake Model [2]. This group has adopted an approach similar to the WGCEP with the intention of creating a global, primarily fault-based model. They have developed a similar model called the "OpenQuake" model that is based on the OpenSHA model of Field et al. (2009).

Another type of earthquake forecast model is a "smoothed seismicity" model in which earthquake seismicity is smoothed in space to produce probabilities for spatial earthquake location occurrence (Rundle et al., 1997; 2002, 2003; Shcherbakov et al., 2005). Applications of this idea are the ETAS and BASS models that assume large earthquake probabilities are a function of the rate of small earthquake activity. Examples of these types of models are the ETAS (2005, 2006, 2007), BASS (Holliday et al., 2005, 2006, 2007) and STEP (Gerstenberger et al., 2005) models, which are statistical models utilizing the Omori and Gutenberg-Richter laws. Rundle et al. (2011) have shown that this assumption is difficult to substantiate with standard tests of forecast accuracy and reliability.

In our previous paper (Rundle et al., 2012) we adopted the basic theoretical framework of repairable systems analysis and lifetime distributions (Xie and Lai, 1996; also ref [3]). We regard earthquake faults as components of a repairable system that is subject to continual stressing. These systems fail but are repairable in the sense that faults heal, only to fail again in the future due to the repeated increase of plate tectonic stresses. We also explicitly assume that earthquakes in a region are correlated with a correlation length $\xi$ as has been seen in observations (Jaume and Sykes, 1999; Jaume, 2000; Hainzl et al., 2000; Zoller et al., 2001). Within this framework, we must compute expressions for the lifetime of the components. Once this is done, we can compute the earthquake (failure) probability of seismically active regions.

**1.2 "Forecast" vs. "Prediction".** Although dictionaries often list these terms as synonyms, we distinguish between a "forecast" and a "prediction" with the following terminology:

o  A "prediction" is a deterministic statement about a future event that can be validated or falsified with a single observation.
o  A "forecast" is a probabilistic statement about a future event that requires multiple observations to establish a confidence level.

In what follows below, we consider only earthquake forecasts. We do not focus on the question of precursors (e.g., Mogi, 1969; Kanamori, 1981; Bufe and Varnes, 1993; Wyss et al., 1996; Bowman and Sammis, 2004; Yen et al., 2006; Huang, 2006; Kossobokov, 2006; Mignan and Giovambattista, 2008; Huang, 2008; Hardebeck et al., 2008; Shearer and Lin, 2009; Greenhough et al., 2009; Rundle et al., 2011). Rather, we focus on the idea of smoothed seismicity models (Tiampo et al., 2002; Rundle et al., 2002; Chen et al., 2005; Chen and Wu, 2006; Kawamura et al., 2013).

**1.3 Synopsis of Method.** The forecast localization method (Rundle et al., 2012) we describe below can be regarded as a Natural Time Weibull (NTW) space-time perturbation on a spatial Relative Intensity (RI) forecast combined with a temporal time-averaged or "Poisson" forecast (Rundle et al., 2002, 2003; Holliday et al, 2005, 2006, 2007).

- We start with a square "pixel" located at $x_i$ and consider a time $t$, around which we construct a large circle of radius $R(x_i,t)$ according to rules described below (Figure 1). The circle of radius $R(x_i,t)$ corresponds to the regional area as described in Rundle et al. (2012).
- The RI piece of the forecast probability for the pixel at $x_i$ and at time $t$ is implemented by a factor $\rho(x_i,t)$ in equation (16) below. This factor allocates forecast weight to the location at $x_i$ in proportion to the level of historic small earthquake activity there in relation to the total small earthquake historic activity within $R(x_i,t)$.
- The time-averaged ("Poisson") large earthquake rate of activity is given by a factor $\Omega(x_i)$. The average number of large earthquakes over the future (forecast) time interval $\Delta t$ is then $n_L = \Omega_i \Delta t$ within the circle of radius $R(x_i,t)$.

The perturbation to the RI-Poisson forecast is $\Psi_i(x_i,t,\Delta t)$ and is defined in equation (18) below. The overall lifetime exponent $h_i[t,\Delta t]$ at the pixel at $x_i$ and at time $t$ is given by:

$$h_i[t,\Delta t] \quad = \quad \text{[RI factor]}$$
$$\text{x [Number Large Earthquakes in } \Delta t\text{]}$$
$$\text{x [ NTW perturbation ]} \qquad (1)$$

The probability of a large earthquake occurring on one of a group of pixels is then:

$$P(\Delta t \mid t) = 1 - \exp\left\{-\sum_i h_i[t, \Delta t]\right\} \qquad (2)$$

The method of construction of the NTW perturbations explicitly assumes that the seismic activity is spatially correlated across the pixels with a correlation length $\xi$.

## 2. Natural Time Weibull Forecast Method

There is a need for a forecast method that accounts both for the current rate of activity, as well as the time since the last major earthquake. This is the natural time Weibull Method ("NTW": Rundle et al., 2012). The NTW method computes conditional (Bayesian) probabilities based on "filling in" the Gutenberg-Richter magnitude-frequency relation:

$$f_\mu = 10^a \, 10^{-bm} \qquad (3)$$

where $f_\mu$ is the frequency (e.g., number per year) of events having magnitude larger than $m$, and $a$ and $b$ are constants (Scholz, 2002).

The NTW method is a "quake count" method. For example, suppose that the last large earthquake was an $m_L = 6$ event, and suppose further example that $b = 1$. The NTW method operates as follows:

o With $b = 1$ the GR relation implies that for every large earthquake with magnitude $m_L$, there are an average $N = 10^{b(m_L - m_S)}$ smaller events with magnitude $m_S$. For $m_L \geq 6$ and $b=1$, there are 1000 $3 \leq m_S < 6$ small earthquakes.
o If 1000 small earthquakes $3 \leq m_S < 6$ earthquakes have occurred since the last large earthquake, the stable GR relation implies that another $m_L \geq 6$ earthquake is required to occur in the relatively near future if the observed GR relation is to continue to be valid.

- o  Weibull statistics are then used to convert the quake count into a probability. We note that Weibull statistics are one of the most common statistical distributions used to describe lifetime and failure (Ebeling, 1997; [3])

Another way to describe this idea is that a system characterized by a statistically stable scaling distribution may temporarily develop a deficiency of large events, since the small events are far more frequent. Eventually, the distribution must be "filled in" by the occurrence of a large event, thereby restoring the statistics for the number of large events relative to the number of small events.

Small events can therefore be used as a kind of "clock" that marks the "natural time" between the large events. This Quake Count method is a variation of the *natural time hypothesis*, and has been discussed in connection with earthquakes (King, 1989; Varotsos et al., 2005; Holliday et al., 2006).

In the previous paper (Rundle et al., 2012), we applied this method, together with backtesting algorithms, to large regions, for example, California and Nevada. In the present paper, we describe a method to spatially localize the NTW forecast to arbitrarily chosen geographic regions.

**2.1 Weibull Probability – Time Domain**

The goal is to compute the probability of a large event occurring within a time interval $\Delta t$ from the present in an arbitrarily defined spatial region. In the calendar time domain, the cumulative Weibull probability for the system to fail at or before time $t$ is defined by:

$$P(t) = 1 - \exp\left\{ -\left(\frac{t}{\tau}\right)^{\kappa} \right\} \quad (4)$$

Here $\tau$ is a time scale, and $\kappa$ is a constant exponent. The Weibull law is used because it is probably the equation most widely used to describe the statistics of failure in engineered systems (Ebeling, 1997; [3]).

The Weibull law has mean $\langle t \rangle = \tau \, \Gamma\left[(\kappa+1)/\kappa\right]$, where $\Gamma[\bullet]$ is the gamma function, and variance $\sigma_t^2 = \tau^2 \left\{ \Gamma\left[(\kappa+2)/\kappa\right] - \left(\Gamma\left[(\kappa+1)/\kappa\right]\right)^2 \right\}$. Note that $\kappa = 1$ corresponds to a Poisson probability.

**2.2 Weibull Probability – Natural Time or Event Count Domain**

On the other hand, in the quake count (QC) or natural time (NT) domain, the Weibull law (Evans et al., 1993; also ref. [3]) is defined by a temporal scale constant $N = \langle N \rangle / \Gamma\left[(\beta+1)/\beta\right]$. Transforming equation (4) to the natural time domain, with parameters $N$, $\beta$, we have:

$$P(n) = 1 - \exp\left\{ -\left(\frac{n}{N}\right)^\beta \right\} \qquad (5)$$

where $n$ is the number of small earthquakes ($m_S$) since the last large earthquake ($m_L$). Equation (5) is the probability of that a large earthquake will occur when $n$ or fewer small earthquakes have occurred.

The goal is to use equation (5) in a forward-looking ("predictive") sense. Therefore we use the conditional (Bayesian) form of the Weibull probability law (5):

$$P(\Delta n \mid n) = 1 - \exp\left\{ -\left(\frac{n+\Delta n}{N}\right)^\beta + \left(\frac{n}{N}\right)^\beta \right\} \qquad (6)$$

Here $P(\Delta n \mid n)$ is a conditional probability. $P(\Delta n \mid n)$ is conditioned on the observation that $n$ small earthquakes $m_S$ have occurred since the last large earthquake having magnitude larger than $m_L$. $P(\Delta n \mid n)$ is the probability that the next large earthquake will occur when an additional $\Delta n$ or fewer small earthquakes have occurred.

As discussed in our previous paper Rundle et al. (2012), an important problem is to relate the natural time interval $\Delta n$ to a calendar time interval $\Delta t$. The most logical assumption for small time intervals $\Delta t$ is to set:

$$\Delta n \approx \omega \, \Delta t \qquad (7)$$

where $\omega$ is the Poisson rate of small earthquakes $m_S$. Equation (7) is the (unconditional) maximum likelihood estimate for the number of small earthquakes occurring during a time interval $\Delta t$ under a Poisson or Gaussian assumption for earthquake probabilities in time (Bevington and Robinson, 1992).

Combining (6) and (7) we find:

$$P(\Delta n \mid n) = 1 - \exp\left\{-\left(\frac{n + \omega \, \Delta t}{N}\right)^\beta + \left(\frac{n}{N}\right)^\beta\right\} \qquad (8)$$

All of the quantities in (8) are observable, an important qualification on the method.

Using equation (8), Rundle et al. (2012) showed that validated probabilities for large, fixed geographical regions could be computed. Regions were chosen for the validations that had at least 6 to 8 large earthquakes in the catalog since 1980, and data for all events were downloaded from the ANSS catalog [4].

For California, the catalog is complete down to about $m_S \geq 3.0$, whereas for the world, the completeness level is generally about $m_S \geq 4.5$ back to about 1980, although in some locations, completeness level can be larger, perhaps $m_S \geq 5.0$. Magnitudes used for "large" earthquakes were California, $m_L \geq 6.0$, and for the world, $m_L \geq 7.0$.

Probabilities were validated by means of the Reliability/Attributes test as well as the Receiver Operating Characteristic test. It was found that in general, $\beta \approx 1.4$. Because the California data is the best in the world, we set $\beta = 1.4$ globally. Testing of this assumption will be ongoing indefinitely (Rundle et al., 2013). Results of these calculations, the time series of probability for California, and Reliability/Attributes and ROC tests are shown in Figures 2-4 and Table 1.

There is a distinct need to develop a method for computing conditional probabilities for arbitrary geographic regions, and this is the problem we address now.

## 3. Localizing NTW Probabilities

Figure 4 illustrates the basic approach. We first consider the question of computing an exceedence probability for large events, i.e., the probability that the next large earthquake $m \geq m_L$ will occur within a time $\Delta t$ from now, conditioned on the previous occurrence of a large earthquake $m \geq m_L$ in the region.

**3.1 Definitions and Components.** The basic ingredients of the localization method are as follows:

o We consider a world-wide earthquake catalog, such as the ANSS catalog, beginning at a time $t_0$, and continuing up to a present time $t$.
o We partition ("tile") the world into 6.48x10$^6$ cells or *pixels* of size 0.1° x 0.1° (we neglect the cells within 10° latitude of the north and south poles).
o At each of the pixels used world-wide ("$i^{th}$ pixel"), extend ("grow") a circle outwards until it contains at least 300 small earthquakes and at least 5 large earthquakes having the target magnitude $m_L$ (Figure 1). Radius of the circle is denoted by $R(x_i, t)$. This region will be used to determine the time averaged ("Poisson") rate $\Omega_i(m_L)$ of earthquakes having magnitude $m \geq m_L$.

- For a seismically active location, $R(x_i,t)$ can be small, of the order of kilometers or tens of kilometers. For less active regions, $R(x_i,t)$ is limited by assumption to radii of a maximum of 1800 km.
- Since small earthquake seismicity tends to be highly clustered in space, it is not generally possible to use the same number of small earthquakes for each pixel.
- The pixel that is at the center of the large circle will be termed the "central pixel"
- The most recent of these large earthquakes will be used as the reference time from which to begin counting small earthquakes.
- The fraction of small earthquakes $m \geq m_S$ contained within the large circle, occurring in the central pixel (pixel $i$) between $(t_0,t)$ will be referred to as $\rho(x_i,t) \equiv \rho_i(t)$.
- The Poisson rate of small earthquakes $m \geq m_S$ within the large circle centered on pixel $i$ is denoted by $\omega_i$. Note that the Poisson rate is the number of such small earthquakes over the long time interval covered by the catalog. In this case, the time interval is from 1980 up to the present day.
- The Gutenberg-Richter $b$-value of the small earthquakes $m \geq m_S$ occurring within the large circle is denoted by $b_i$
- The number of small earthquakes since the last large earthquake within the large circle will be denoted by $n_i(t)$.
- Within the large circle centered on pixel $i$, the number $N_i$ of small earthquakes having magnitude $m_S \leq m < m_L$ corresponding to 1 large earthquake of magnitude $m_L$ is $N_i = 10^{b_i(m_L - m_S)}$.
- Given the Gutenberg-Richter $b$-value of the small earthquakes $m \geq m_S$, we note that the expected Poisson rate $\Omega_i$ for the large earthquakes $m \geq m_L$ within the large circle is:

$$\Omega_i = \frac{\omega_i}{N_i} = \omega_i 10^{-b_i(m_L - m_S)} \tag{9}$$

Note that in many parts of the world, catalog completeness issues may imply that $\Omega_i$ can be more accurately determined than $\omega_i$, while also noting that $b$-values can be determined relatively well from recent magnitude-frequency data (see the discussion in section 3.4).

- As a result, a better strategy is to use the observed rate of large earthquakes directly and extrapolate these to small earthquake rates if needed.
- We note that if the previous conditions of 5 large earthquakes within 1800 km radius is not possible, then determine the largest value of $m_L$ for which it is possible and its associated circle radius $R(x_i,t)$. Then use the GR relation to extrapolate:

$$\Omega_i(m) = \Omega_i(m_L)\ 10^{b(m_L - m)}, \qquad m > m_L \qquad (10)$$

- Given an optimal value of exponent $\beta$, we compute the lifetime factor (exponent of the probability) by the method of Rundle et al. (2012):

$$H_i[n_i,\ \Omega_i \Delta t, m \geq m_L] \equiv \left(\frac{n_i}{N_i} + \Omega_i \Delta t\right)^\beta - \left(\frac{n_i}{N_i}\right)^\beta \qquad (11)$$

**3.2 Effects of More Distant Earthquakes.** There is a problem which is involved with accounting for the effects of more distant earthquakes, both small and large, outside the circle of radius $R(x_i,t)$. In models of the earthquake process published in the recent past (Klein et al., 2000), it has been pointed out that earthquakes are spatially correlated with a correlation length roughly equal to the size of the largest earthquake in the region. In the present work, we assume that this distance is about 400 km, which is a typical source dimension for large and great earthquakes.

The method by which these more distant earthquakes are incorporated into the algorithm is by the use of the correlation function. We adopt the simplest method, noting that the correlation function $C(r)$ for a mean field system in $d = 2$ dimensions is (Goldenfeld, 1992):

$$C(r) = e^{-r/\xi} \qquad (12)$$

where $r$ is radial distance and $\xi$ is the correlation length. Note that $C(r) \leq 1$.

We use the $d = 2$ dimensional correlation function since the distances between the large earthquakes are usually to be significantly than their depths, which are often crustal depths < 30 km or so. In other words, if depth << $\xi$, then the $d = 3$ dimensional correlation function is well approximated by (12). In addition, depth data in the global catalogs are less reliable than latitude-longitude position. Future work (Rundle et al., 2013) will explore the effects of using the $d = 3$ dimensional correlation function.

We use (12) in the following way. Note that for small earthquakes $m_L > m \geq m_S$, if the location $x_S$ of the small earthquake is inside the circle of radius $R(x_i,t)$, i.e., $r \equiv |x_S - x_i| > R(x_i,t)$, we increment the sum $n_i(t)$ by:

$$n_i(t) \rightarrow n_i(t) + 1 \qquad \text{for } r \leq R(x_i,t) \qquad (13)$$

However, for small events outside the circle of radius $R(x_i,t)$, i.e., $r \geq R(x_i,t)$, we increment the sum $n_i(t)$ by:

$$n_i(t) \rightarrow n_i(t) + e^{-(r-R)/\xi} \qquad \text{for } r \geq R(x_i,t) \qquad (14)$$

In words, a more distant small earthquake increases the event count, but not by as much as such a small earthquake within the range of correlation.

Likewise, for large earthquakes for which $m \geq m_L$, one might expect that the overall total stress level in the earth has declined, and that such an event decreases the likelihood of another such event. Within the radial distance $r \leq R(x_i,t)$:

$$n_i(t) \rightarrow 0 \qquad \text{for } r \leq R(x_i,t) \qquad (15)$$

Outside the radial distance $r \geq R(x_i,t)$, we again use (12) to incorporate this aspect of the model:

$$n_i(t) \rightarrow n_i(t) \times (1 - e^{-(r-R)/\xi}) \qquad \text{for } r \geq R(x_i,t) \qquad (16)$$

In words, a large distant earthquake decreases the small $n_i(t)$ event count, but not by as much as a large event within the range $r \leq R(x_i,t)$ for which $n_i(t) \to 0$ following the large earthquake.

**3.3 Spatial and Temporal Correction Factors.** $H_i[n_i, \Omega_i \Delta t, m \geq m_L]$ must be computed from some initial time $t_0$ up to the current time $t$ using only prior catalog data. This means that all quantities appearing in $H_i[n_i, \Omega_i \Delta t, m \geq m_L]$, including the Poisson rate, are implicitly functions of time $t$.

We must account for the idea that only a fraction $\rho_i$ of the small earthquake activity within the large circle is associated with the spatial pixel of interest. In addition, we should also recognize that the lifetime factor $H_i[n_i, \Omega_i \Delta t, m \geq m_L]$ must be consistent with the constraint of an observable time averaged ("Poisson") rate of large earthquakes within the large circle of radius $R(x_i,t)$. We therefore determine a time-dependent correction factor $f_i(t)$ for each pixel.

We first define $n_L \equiv \Omega_i \Delta t$, the average number of large events within the forecast time interval $\Delta t$ of the target magnitude $m \geq m_L$ within the circle of radius $R(x_i,t)$. We note that $n_L$ can be determined either by extrapolation from small magnitude earthquakes using the Gutenberg-Richter relation, or directly from the large magnitude events themselves. We have found that in general, determining $n_L$ directly from the number of large magnitude events seems to be the best approach. This is because the catalogs are more likely to be complete at the large magnitude level than at the small magnitude level.

Then the normalization is defined to be:

$$n_L = f_i(t) \frac{1}{(t-t_0)} \int_{t_0}^{t} H_i[n_i(t'), \Omega_i(t')\Delta t, m \geq m_L] \, dt' \qquad (17)$$

so that:

$$f_i(t) = \frac{n_L}{\frac{1}{(t-t_0)} \int_{t_0}^{t} H_i[n_i(t'),\ \Omega_i(t')\Delta t, m \geq m_L]\ dt'} \quad (18)$$

We define:

$$h_i[t,\Delta t] = \rho_i(t)\ f_i(t)\ H[n_i,\ \Omega_i \Delta t, m \geq m_L] \quad (19)$$

The factor $\rho_i(t)$ allocates a fraction of the lifetime factor to the pixel $x_i$ corresponding to the historical fraction $\rho_i(t)$ of the small earthquake seismicity that occurs within the circle of radius $R(x_i,t)$ centered at $x_i$. In other words, $\rho_i(t)$ is the total number of historical small events in the central pixel at time $t$ divided by the total number of historical small events within the circle of radius $R(x_i,t)$.

The factor $f_i(t)$ makes the lifetime factor consistent with the long term average ("Poisson") rate of large earthquake activity in the region within $R(x_i,t)$. We may alternately write equation (18) as:

$$h_i[t,\Delta t] = \rho_i(t)\ n_L(x_i)\ \Psi(x_i,t,\Delta t) \quad (20)$$

where:

$$\Psi(x_i,t,\Delta t) = \frac{\int_{t_0}^{t} H_i[n_i(t'),\ \Omega_i(t')\Delta t, m \geq m_L]\ dt'}{\frac{1}{(t-t_0)} \int_{t_0}^{t} H_i[n_i(t'),\ \Omega_i(t')\Delta t, m \geq m_L]\ dt'} \quad (21)$$

Equation (21) is the NTW contribution to the overall lifetime factor $h_i[t,\Delta t]$ and may be regarded as a time-dependent perturbation on the time averaged rate $\Omega_i$ of large earthquakes within $R(x_i,t)$.

The NTW probability for a large earthquake occurring at the pixel located at $x_i$ within $\Delta t$ after time $t$ is then:

$$P(\Delta t \mid x_i, t) = 1 - \exp\{-h_i(t, \Delta t)\} \qquad (22)$$

By construction, the probability for a large earthquake occurring in any of the subset $\{x_i\}$ of pixels is:

$$P(\Delta t \mid x_i, t) = 1 - \exp\left\{-\sum_i h_i(t, \Delta t)\right\} \qquad (23)$$

The assumptions involved in going from equation (22) to (23) are that:

- The circle of radius $R(x_i, t)$ is large enough so that $h_i(t, \Delta t)$ does not depend sensitively on details of its size. We have conducted a series of trials with varying $R(x_i, t)$ that lend support to this assumption.
- The lifetime factor of large earthquake occurrence within the subset of pixels $\{x_i\}$ is proportional to the fraction of small earthquakes that occur within $\{x_i\}$ relative to the total number that occur within $R(x_i, t)$. This latter assumption finds support from the RELM test of earthquake occurrence that has been reported elsewhere

Note that the model explicitly assumes that seismic activity is spatially correlated with a correlation length $\xi$, and that the contribution to the exponent from the pixel at $x_i$ is $h_i(t, \Delta t)$.

**3.4 Global Data Issues.** Our goal is to produce a method that is valid world-wide. As a result, a major constraint on the efficacy of the method is variation in the quality of the data with location. For example, the quality of the data is highest in the United States, specifically California. Here the data catalogs are generally complete (at least in southern California) back to 1932 down to magnitudes of about $m \geq 3.0$. However, even though these early events may be included in the catalog, details of their source parameters such as locations magnitude determinations have improved substantially in the modern era (since about 1990).

By contrast, the global digital network has only been in place for a decade or two, and has been upgraded with more stations over the years since its establishment in 1986 [5]. A number of global locations are poorly covered even today by these stations, particularly in areas of Asia including China, India and the middle east, such as Pakistan, Iran, and Iraq. Since our method uses data back to 1980 to determine event counts, this lack of completeness in the small events of the global catalog is a serious problem. A manifestation of this problem is that average ("Poisson") rates of large event activity that is extrapolated from historic small earthquake rates is generally too small in many regions of southern Asia and the middle east. However, global events larger than $m \geq 6$, have been generally observed during the last 30 years and are present in the catalog.

The computation of large earthquake average ("Poisson") rates in these deficient regions must be improved. One strategy to do so is to draw circles having larger radii $R(x_i,t)$ that contain a sample of large magnitude earthquakes $m \geq m_L$. Typically we prefer to use at least 5 large earthquakes in these cases. The most recent of these 5 large earthquakes remains the reference event for counting small events.

We note that Gutenberg-Richter plots, which are often used to judge the completeness of a catalog, are often not useful for this purpose in our application. If most of the events in the catalog come from the last 5 years or so, which is often the case, a log frequency-magnitude GR diagram may appear linear due to the recent events. However, it could still have major data gaps at earlier time periods which precludes backtesting or other "retrocasting" that requires small event data at these earlier time periods.

## 4. Forecast Validation and Verification.

In a previous paper we discussed the issue of backtesting, validation and verification. Since we are dealing with a forecast here, we use validation and verification tests that have a substantial provenance, as discussed previously (Green and Swets, 1966; Winkler and Murphy, 1968; Murphy, 1973; Mason, 1982; Murphy and Daan, 1985; Hsu and Murphy, 1986; Murphy and Winkler, 1987; Murphy, 1988; Kharin and Zwiers, 2003;

Mason, 2004; Atger, 2004; Joliffe and Stephenson, 2005; Casati et al., 2008; Rundle et al., 2012; see also note [6]).

There are two important parameters to be determined by backtesting, specifically the exponent $\beta$ and the correlation length $\xi$. In Rundle et al. (2012) we also discussed another parameter $\alpha$ multiplying the average ("Poisson") rate, but the requirement that the lifetime factor be regarded as a perturbation on the time averaged ("Poisson") rate removes this parameter from consideration. In what follows, we also choose $\xi = 400$ km uniformly. Theoretical considerations (Klein et al., 2000) imply that $\xi$ should be approximately the source dimension of the largest earthquake in the region. For California, this linear dimension is indeed about 400 km, but for other parts of the world, particularly subduction zones, this value may be different. However, for purposes of simplicity, we adopt the convention that $\xi = 400$ km worldwide.

For now this leaves us the task of determining $\beta$ from backtests. In Rundle et al. (2012) we used Reliability/Attributes (R/A) tests and Receiver Operating Characteristic (ROC) tests to determine a value $\beta \approx 1.4$ for both the fixed California-Nevada region, as well as the Japan region. In the R/A test, one uses the Briar score, together with measures of reliability error, resolution and skill. In the ROC test, one uses the ROC curve as well as integrals such as the Area Skill Score. Bootstrap error analyses can be used to estimate confidence levels (Efron and Tibshirani, 1993).

Results of these are presented in Table 1 for the California-Nevada region, for two time periods. The first time period is for 1980-2012.95 (the most recent $m_L \geq 6$ earthquake was about 150 km SW of San Diego on December 16, 2012). The second time period is for 1995-2012.95. As explained in Rundle et al. (2012), we terminate the bootstrap analysis on the last large earthquake to occur, since any further forecast data will necessarily be counted as a false alarm by the testing procedures. The two initial dates account for, in the first case (1980), the onset of somewhat reliable analog data. The second intial date (1995) corresponds to the date at which the network in the California-Nevada was entirely digital, with uniform reporting standards.

As can be seen in the time series in Figure 2, there occurrence of the $m_L \geq 6$ occurred at values of maximum values of probability of about 50% until 1995, after which the maximum value of probability is observed to be about 60%. The smaller values of probability may result from either incompleteness of the catalog (missing events) at the small event magnitudes. Another possibility is the underestimation of the magnitude of the small events. In either case, $n_i(t)$ would tend to be too small at the time the large earthquake occurred, therefore leading to a lower value of peak probability.

In further spot testing at various locations around the world, we continue to find that $\beta \approx 1.4$ represents a broad, flat minimum in the Briar reliability error. We will defer further consideration of these continuing and ongoing validation/verification activities to future publications (Rundle et al., 2013).

## 5. User Interface

One of the challenges (Rundle et al., 2013) in forecasting research is to present the information in usable form. To date, forecast information is typically presented in journal articles that are appropriate for archival purposes. However, earthquake forecasting information is dynamic, and requires a platform that allows computation, updating and independent analysis in real time. For that reason, we have encoded the methods described here into an open access online platforms [7-9]. A User Interface (UI) includes a toolset that allows retrieval of forecast information world-wide ("viewer tool") for display onscreen via the standard web browsers.

The general workflow is as follows. Every evening at about midnight east coast time, catalog feeds are downloaded and combined to form the input data. These feeds are primarily the ANSS catalog and the USGS 30-day real time feed. Other data feeds may be used as appropriate. Since the ANSS catalog is not always updated in real time, and since our goal is to provide a real time forecast, we need to ensure that the catalog is in fact updated daily. When combining the catalogs, event IDs are checked to eliminate the possibility of including multiple listings of the same event.

The methods described in the foregoing discussion are then applied to compute the lifetime factor $h_i[t,\Delta t]$ at pixels of $0.1^o$ as discussed above following the download of the data. These lifetime factors are then assigned as tags to each screen pixel, and also used to construct a KML file for displaying the forecast on screen. Since the NTW lifetime factor is nonzero only at pixels where seismicity exists, about 4% of the earth's surface, the NTW forecast by itself has sharp spatial boundaries. However, some of this sharpness can be due to errors in location of the catalog data.

As a result, we invoke a spatial smoothing of the lifetime factors, typically a Gaussian smoothing over radial distances of about $0.2^o$. In addition, where recent activity has been low, but previous activity has been high, the average "Poisson" rate is not negligible. The forecast that is finally displayed on the screen is therefore a combination, or ensemble forecast consisting of 80% NTW and 20% smoothed BASS (ETAS) forecast (Holliday et al., 2007). We have found that this combination of validated forecasts provides adequate spatial smoothing consistent with uncertainty in global earthquake locations, while at the same time providing enhanced aftershock probability in a region while maintaining the great majority of the NTW forecast probability.

A second important issues is the need for a rapid response time in a web-based forecast. Users will typically wait for periods of only seconds for a response from a query. While this is not important for research data, it is nevertheless important if the idea is to increase practical utility of the data. For that reason, we have made several choices in the architecture of the site that emphasize speed of response.

An example is the forecast timeseries tool. This tool must parse many gigabytes of data to return a timeseries for the selected region. In addition, most web browsers have a relatively small time out setting. So if the results are to be delivered to users via a browser, the amount of data delivered cannot be unlimited. As a result, we limit the timeseries to a past interval of only 5 years rather than the 32+ years of total data. In addition, the selected region is also generally limited to about $3 \times 10^4$ km$^2$ for the same reason, unless the user chooses to reset the wait time on their browser.

To use these tools, the user navigates to the page [9]. The user can then invoke selection tools (radio buttons) to locate a circular region anywhere on the planet with arbitrary radius, or a polygon selection allowing regions of arbitrary shape. The system then sums the lifetime exponents for all of the pixels in the selection region, and computes and displays the probabilities in tabular form. The table appears in the lower left corner of the viewer page listing the forecast probabilities for various time periods into the future (1 month, 1 year, 3 years) and for various magnitude levels ($m \geq 5, m \geq 6, m \geq 7, m \geq 8$).

If the selected region is small enough, one can also display the forecast probability timeseries by clicking on the button labeled "Forecast Timeseries". If the selected region is too large (radius > 100 km or so), the browser will typically time out after waiting 60-120 seconds and a "no data" warning will be displayed. It is possible to increase the time out period for some browsers, but not for all. These practical considerations do not, of course, apply for research or offline forecast computations.

## 6. Results and Discussion

Figures 2-4 show updated plots of regional forecasts for California-Nevada and for Japan, together with results from the R/A and ROC backtests. Earlier versions of these plots appeared in our previous paper (Rundle et al., 2012). Note that the backtests are carried out for time periods from 1980 to 2012.95 = December 16, 2012, the time of occurrence of the latest earthquake. As discussed, this is because the time period following the latest large earthquake will always be counted as a false alarm, since no large earthquake has as yet occurred.

Screen shots from [7,9] are shown in Figures 5-8. Figures 5 and 7 are maps of California and Japan, respectively, displaying selection circles, together with contours of the NTW forecast. The circular selection region in California has a radius of 100 km, while the selection region for Japan has a radius of 150 km. For California, we show in Figure 6 the time-dependent nature of the forecast probabilities for $m_L \geq 6.0$ events in the selection circle of Figure 5 over the course of the 5+ years since January 1, 2008.

One can observe sudden decreases in the forecast due to the occurrence of nearby earthquakes having $m_L \geq 6.0$. These events outside the circular selection region influence the probability of large events inside the selection region via the influence of the exponential term in equation (16). Sudden increases in probability are generally due to aftershocks of the large nearby earthquakes occurring either in the circular selection region itself, or farther away, influencing the probability via the exponential term in equation (14). Figure 8 shows a similar timeseries plot of forecast probability for the circular selection region centered on Tokyo, Japan.

A User Interface of the type described here provides a means of discovery that has not previously been available. Referring to the Japan example, we can see that computed forecast probabilities for a large or great earthquake in the next 1-3 years are quite high, as can be seen from the viewer site, and are shown in Table 3, even for events having $m \geq 8.0$. We may ask whether these probabilities are reasonable, inasmuch as the great $m = 9.1$ earthquake on March 11, 2011 in Tohoku might be presumed to have relieved the regional stress level and therefore decreased the likelihood of another great earthquake.

To answer this question, recall that the NTW method simply presumes that over time, the Gutenberg-Richter relation will be complete at all magnitude levels. If there has been a large number of small earthquakes, and a relative deficiency of large earthquakes, then it seems reasonable to expect that this deficiency will be filled in. When the next large or great earthquake occurs is determined by the accumulation of future earthquakes in the selected region. We estimate the accumulation rate by the Poisson rate of small earthquakes and scale this small earthquake rate to the expected accumulation rate of large earthquakes to convert from natural or event count time to calendar time.

To make this point more clearly about Japan, Figure 9 shows the Gutenberg-Richter relation for the 150 km circular region around Tokyo, for all events just after the March 11, 2011 earthquake up to the present (June 12, 2013). It can be seen that the number of earthquakes larger than $m_L \geq 6.0$ falls below the scaling line that has been fit to the small magnitude earthquake distribution. There is an obvious deficiency of earthquakes $m_L \geq 6.0$, meaning that larger earthquakes are needed to complete the linearity of the Gutenberg-Richter relation.

Finally, we comment on the debate regarding whether great earthquakes trigger other great earthquakes globally, possibly leading to the apparent clustering of such great events that has recently been observed since about 2004. This can be interpreted as two possibly unrelated phenomena, causality and correlation. On the one hand, there is the problem of whether great earthquakes trigger, or cause other great earthquakes by some as yet poorly understood physics. On the other hand, there is the problem of whether the clustering represents the synchronization of some global processes, or whether it arises from purely random events.

Shearer and Stark (2011) have discussed these ideas and conclude that the statistical evidence for such triggering or even anomalous clustering is weak. Van der Elst et al. (2013) examine additional data on great earthquakes from 1998-2011 and conclude that large earthquakes do not appear to cause an increase in rate of other large earthquakes elsewhere on the planet. They also estimate that a smaller catalog completeness level down to $m \geq 2.0$ is necessary for the International Seismic Center (ISC) catalog to resolve the question of whether regional triggering exists.

All of these conclusions are based on statistical analysis of the available catalog data, which is necessarily incomplete in important respects. The basic problem is the fact that the instrumental record is very short compared to the average recurrence time of such great earthquakes, so we are confined to the statistics of small samples.

The debate about triggering is a conversation about whether one great earthquake deterministically causes another great earthquake elsewhere on the planet. The NTW model only considers the accumulation of events in a region making up the Gutenberg-Richter relation, and does not assume any form of triggering. If no increase in small earthquake rates from distant great earthquakes occurs, then there will be no change in the Gutenberg-Richter relation and no local change in the probability of great earthquakes. NTW therefore implies nothing about causality or synchronization, it is only based on statistically completing the Gutenberg-Richter relation over time.

**Acknowledgements.** Research by JBR and JRH was performed with funding from NASA NNX12AM22G to the University of California, Davis.

**Appendix**

**Probability Inversions and Magnitude Bands**. We have found that in many locations in the world, problems with data quality lead to spurious results if a simple exceedance are computed. A manifestation of this is that apparent forecast "inversions" can occur, whereby forecast probabilities for $m_L \geq 7$ events may be larger than forecast probabilities for $m_L \geq 6$, for example, obviously a non-physical result. One strategy to deal with this problem is to compute a forecast in a band of large earthquake magnitudes $m_L \in [M_1, M_2]$, where $M_2 > M_1$, rather than computing an exceedance forecast for $m_L \geq M$.

To compute a forecast for a magnitude band $[M_1, M_2]$, we observe that the numbers $N_1$ and $N_2$ of small earthquakes $m$ corresponding to larger earthquakes $M_1$ and $M_2$ are:

$$N_1 = 10^{b(M_1 - m)}$$

$$N_2 = 10^{b(M_2 - m)} \tag{22}$$

The difference between these two numbers:

$$N^* = N_2 - N_1 = 10^{b(M_2 - m)} - 10^{b(M_1 - m)}$$

$$= 10^{-bm}\left[10^{bM_2} - 10^{bM_1}\right] \tag{23}$$

is a characteristic (reference) number of small earthquakes associated with this band of large magnitudes.

If we do some simple manipulations, we can put this into the desired form. First define:

$$10^{bM^*} = 10^{bM_2} - 10^{bM_1} \tag{24}$$

Then we see from (9) that:

$$N^* = 10^{b(M^* - m)} \qquad (25)$$

It is then easy to find that:

$$M^* = M_1 + (1/b)\, Log_{10}(10^{b\,\Delta M} - 1) \qquad (26)$$

The number $N^*$ and the large magnitude $M^*$ are then characteristic of the band from $M_1$ to $M_2$, where $\Delta M = M_2 - M_1$.

To implement this additional feature of forecasting within a magnitude band centered on the magnitude $m_j$, we use the definition as in (9):

$$H_i[n_i,\, \omega_i \Delta t, m_j] \equiv \left(\frac{n_i + \omega_i \Delta t}{N_i^*}\right)^\beta - \left(\frac{n_i}{N_i^*}\right)^\beta \qquad (27)$$

and proceed as previously. To compute an exceedance forecast for $m \geq m_L$ rather than within a magnitude band $m_j \in [M_j, M_{j+1}]$, we sum over all bands for which $m_j \geq m_L$:

$$H_i[n_i,\, \omega_i \Delta t, m \geq m_L] \equiv \sum_j H_i[n_i,\, \omega_i \Delta t, m_j] \qquad (28)$$

Since all of the $H_i[n_i,\, \omega_i \Delta, m_j] \geq 0$, we are guaranteed that there will be no forecast "inversions" of the type described above.

**Notes.**

[1] www.wgcep.org ; http://pubs.usgs.gov/of/2007/1437/

[2] www.globalearthquakemodel.org

[3] *NIST/SEMATECH e-Handbook of Statistical Methods*,
   http://www.itl.nist.gov/div898/handbook/
   Weibull:  http://www.itl.nist.gov/div898/handbook/apr/section1/apr162.htm

[4] NCEDC  http://www.ncedc.org/cnss/catalog-search.html

[5] http://pubs.usgs.gov/fs/2011/3021/

[6] CAWCR,  http://www.cawcr.gov.au/projects/verification/

[7] www.openhazards.com

[8] www.quakesim.org

[9]  www.openhazards.com/viewer

**References.**

Atger, F., Relative impact of model quality and ensemble deficiencies on the performance of ensemble based probabilistic forecasts evaluated through the Brier score, *Nonlin. Proc. Geophys.*, 11, 399-409 (2004).

Bowman, D.D. and C.G. Sammis, Intermittent criticality and the Gutenberg-Richter distribution, Pure Appl. Geophys., 161,1945-1956 (2004).

Bufe, C.G. and D.J. Varnes, Predictive modeling of the seismic cycle in the San Francisco Bay region, *J. Geophys. Res.*, 98, 9871-9883 (1993).

Bevington, P.R. and D.K. Robinson, *Data Reduction and Error Analysis for the Physical Sciences*, McGraw-Hill, New York (1992)

Casati, B., L.J. Wilson, D.B. Stephenson, P. Nurmi, A. Ghelli, M. Pocernich, U. Damrath, E.E. Ebert, B.G. Brown and S. Mason, Forecast verification: current status and future directions, *Meteorol. Appl.*, 15, 3-18 (2008)

Chen, CC, and Wu, YX, An improved region-time-length algorithm applied to the 1999 Chi-Chi, Taiwan earthquake, *Geophys. J. Int.* 166, 1144-7 (2006)

Chen, CC, Rundle, J.B., Holliday, J.R., Nanjo, KZ, Turcotte, DL, Li, S-C and Tiampo, KF, The 1999 Chi-Chi, Taiwan, earthquake as a typical example of seismic activation and quiescence,  *Geophys. Res. Lett.,* **32**, L22315 (2005)

Ebeling, Charles E., *An Introduction to Reliability and Maintainability Engineering*,

McGraw-Hill, Boston (1997)

Efron, B. and Tibshirani, R. , *An Introduction to the Bootstrap*, Chapman and Hall/CRC, Boca Raton, FL (1993)

Evans, M., N. Hastings, B. Peacock, *Statistical Distributions*, John Wiley & Sons (1993)

Field, EH et al., Uniform California earthquake rupture forecast, version 2 (UCERF2), *Bull. Seism. Soc. Am.*, **99**, 2053-2107 (2009).

Gerstenberger, MC, S. Weimer, LM Jones and P Reasenberg, Real time forecasts of tomorrow's earthquakes in California, *Nature*, 435, 328-331 (2005)

Goldenfeld, N., *Lectures on Phase Transitions and the Renormalization Group*, Addison Wesley (1992)

Green, D.M. and J.M. Swets, *Signal detection theory and psychophysic,* New York: John Wiley and Sons (1966).

Greenhough, J., A. Bell & I.G. Main (2009). Comment on "Relationship between accelerating seismicity and quiescence, two precursors to large earthquakes" by Arnaud Mignan and Rita Di Giovambattista, Geophys. Res. Lett. 36, L17303, doi:10.1029/2009GL039846

Hainzl, S., G. Zoller, J. Kurths, and J. Zschau, Seismic quiescence as an indicator for large earthquakes in a system of self-organized criticality, *Geophys. Res. Lett.*, 27, 597– 600. (2000)

Hardebeck, J. L., K. R. Felzer, and A. J. Michael, Improved tests reveal that the accelerating moment release hypothesis is statistically insignificant, *J. Geophys. Res.*, 113, B08310, doi:10.1029/2007JB005410. (2008)

Holliday, J.R., K.Z. Nanjo, K.F. Tiampo, J.B. Rundle and D.L. Turcotte, Earthquake forecasting and its verification, *Nonlin. Proc. Geophys.*, 12, 965-977 (2005).

Holliday, J.R., J.B. Rundle, D.L. Turcotte, W. Klein and K.F. Tiampo, Space-time correlation and clustering of major earthquakes *Phys. Rev. Lett.*, **97**, 238501 (2006)

Holliday, J.R., D.L. Turcotte and JB Rundle, BASS, an alternative to ETAS, *Geophys. Res. Lett.*, **34** Art. No. L12303 doi:10.1029/2007GL029696 (2007).

Hsu, W.-R. and A.H. Murphy, The attributes diagram: A geometrical framework for assessing the quality of probability forecasts, *Int. J. Forecasting*, 2, 285-293 (1986)

Huang, Q, Search for reliable precursors: a case study of the seismic quiescence of the 2000 western Tottori prefecture earthquake, *J. Geophys. Res.*, 111, . (2006).

Huang, Q, Seismicity changes prior to the $M_s$8.0 Wenchuan earthquake in Sichuan, China, *Geophys. Res. Lett.*, 35, L23308 (5 pp.) (2008)

Jaume, S. C. and L.R. Sykes, Evolving towards a critical point: a review of accelerating seismic moment/energy release prior to large and great earthquakes, *Pure Appl. Geophys.* 155, 279-305 (1999).

Jaume, S. C., Changes in earthquake size-frequency distributions underlying accelerating seismic moment/energy release, AGU Monograph *Geocomplexity and the Physics of Earthquakes*, ed. by J.B. Rundle, D.L. Turcotte, and W. Klein, p. 199-210 (2000)

Joliffe, IT and DB Stephenson, *Forecast Verification: A Practitioners' Guide in Atmospheric Science*, John Wiley & Sons (2003)

Kanamori, H., The nature of seismicity patterns before large earthquakes, in Ewing, M., ed., Series 4: *Earthquake Prediction - An International Review*, AGU Geophys. Mono.: Washington D.C., p. 1-19 (1981).

Kawamura, M, Y-H Wu, T. Kudo and C.C. Chen Precursory migration of anomalous seismic activity revealed by the Pattern Informatics method: A case study of the 2011 Tohoku earthquake, Japan, *Bull. Seismi. Soc. Am.*, 103, 1171-1180 (2013).

Kharin, V.V. and F.W. Zwiers, On the ROC score of probability forecasts, *J. Climate*, 16, 4145-4150 (2003)

King, G., Variance specification in event count models: From restrictive assumptions to a generalized estimator, *Am. J. Pol. Sci.*, **33**, 762-784 (1989)

Klein, W., M. Anghel, C.D. Ferguson, J.B. Rundle and J.S.S. Martins, Statistical analysis of a model for earthquakes faults with long-range stress transfer, 43-72, in *GeoComplexity and the Physics of Earthquakes*, ed. J.B. Rundle, D.L. Turcotte and W. Klein, American Geophysical Union, Washington, DC (2000).

Kossobokov, VG, Testing earthquake prediction methods: The West Pacific short-term forecast of earthquakes with magnitude Mw > 5.8, *Tectonophysics*, 413, 25-31 (2006)

Mason, I.B., A model for assessments of weather forecasts, *Austral. Met. Mag*, **30**, 291-303 (1982)

Mason, S.J., On using "climatology" as a reference strategy in the Brier and ranked probability skill scores. *Mon. Wea. Rev.*, 1891-1895 (2004).

Matthews, M.V., W.L. Ellsworth and P.A. Reasenberg, A Brownian model for recurrent earthquakes, *Bull. Seism. Soc. Am.*, **92**, 2233-2250 (2002).

Mignan, A., Di Giovambattista, R., Relationship between accelerating seismicity and quiescence, two precursors to large earthquakes, *Geophys. Res. Lett*., 35, L15306 (5 pp.) (2008)

Mogi, K., Some features of recent seismic activity in and near Japan (2), Activity before and after great earthquakes, *Bull. Earthquake Res. Inst. Univ. Tokyo*, 47, 395–417. (1969)

Murphy, A.H and H. Daan, Forecast evaluation, in: A.H. Murphy and R.W. Katz, eds., *Probability, Statistics and Decision Making in the Atmospheric Sciences*, Westview Press, Boulder, CO (1985)

Murphy, A.H. and R.L Winkler, A general framework for forecast verification, *Mon. Weather Rev.*, **115**, 1330-1338 (1987).

Murphy, A.H., A new vector partition of the probability score, *J. Appl. Meteor*., 12, 595-600 (1973).

Murphy, A.H., Skill scores based on the mean square error and their relationships to the correlation coefficient. *Mon. Wea. Rev.*, **116**, 2417-2424 (1988).

Ogata, Y., Synchronous seismicity changes in and around the northern Japan preceding the 2003 Tokachi-oki earthquake of M8.0, *J. Geophys. Res.,* **110**, B08305, (2005)

Ogata, Y., Monitoring of anomaly in the aftershock sequence of the 2005 earthquake of M7.0 off coast of the western Fukuoka, Japan, by the ETAS model, *Geophys. Res. Lett*., 33, L01303 (2006).

Ogata, Y., Seismicity and geodetic anomalies in a wide area preceding the Niigata-Ken-Chuetsu earthquake of 23 October 2004, central Japan, *J. Geophys. Res.*, 112, B10301-1-11 (2007).

Rundle, J.B. et al. (to be published, 2013)

Rundle, J.B., K.F. Tiampo, W. Klein and J.S.S. Martins, *Proc. Nat. Acad. Sci*. USA, **99**, Supplement 1, 2514-2521, (2002)

Rundle, J.B., DL Turcotte, C Sammis, W Klein and R. Shcherbakov, Statistical physics approach to understanding the multiscale dynamics of earthquake fault systems, *Rev. Geophys. Space Phys.*, **41**(4), DOI 10.1029/2003RG000135 (2003).

Rundle, JB, PB Rundle, A Donnellan, D Turcotte, R Shcherbakov, P Li, BD Malamud, LB Grant, GC Fox, D McLeod, G Yakovlev, J Parker, W Klein, KF Tiampo, A simulation-based approach to forecasting the next great San Francisco earthquake, *Proc. Nat. Acad. Sci.*, 102: 15363-15367 (2005) ; published online before print October 11 2005, 10.1073/pnas.0507528102

Rundle, JB, S Gross, W Klein, and DL Turcotte, The statistical mechanics of earthquakes, *Tectonophysics*, *277*, 147-164 (1997).

Rundle, J.B., J.R. Holliday, M Yoder, M. K. Sachs, A. Donnellan, D. L. Turcotte, K. F. Tiampo, W. Klein and L. H. Kellogg, Earthquake precursors: activation or quiescence?, *Geophys J. Int.*, **187**, 225-236 (2011)

Rundle, J.B., J.R. Holliday, W.R. Graves, D.L. Turcotte, K.F. Tiampo and W. Klein, Probabilities for large events in driven threshold systems, *Phys. Rev. E*, **86**, 021106 (2012)

Rundle, J.B. et al., to be published (2013).

Scholz, C.H., *The Mechanics of Earthquakes and Faulting*, Cambridge Univ. Press (2002)

Shcherbakov, R, G Yakovlev, DL Turcotte and JB Rundle, Model for the distribution of aftershock interoccurrence times, *Phys. Rev. Lett.*, *95*, Art. 218501 (2005).

Shearer, P. M., and G. Lin Evidence for Mogi doughnut behavior in seismicity preceding small earthquakes in southern California, *J. Geophys. Res.*, *114*, B01318, doi:10.1029/2008JB005982. (2009)

Shearer, P.M. and P.B. Stark, Global risk of big earthquakes has not recently increased, *Proc. Nat. Acad. Sci.*, **109**, 717-721 (2011).

Tiampo, KF, JB Rundle, S. McGinnis and W. Klein, Pattern dynamics and forecast methods in seismically active regions, *PAGEOPH*, 159, 2429-2467 (2002).

Van der Elst, N.J, E.E. Brodsky, and T. Lay, Remote triggering not evident near epicenters of impending great earthquakes, *Bull. Seism. Soc. Am.*, 103, 1522-1540 (2013). doi: 10.1785/0120120126.

Varotsos, P., N.V. Sarlis, H.K. Tanaka and E.S. Skordas, *Phys. Rev. E*, **71**, 032102 (2005)

Winkler, R.L. and A.H. Murphy, 'Good' probability assessors, *J. Appl. Meteor.*, 7, 751-758 (1968).

Wyss, M., K. Shimazaki, and T. Urabe, Quantitative mapping of a precursory seismic quiescence to the Izu-Oshima 1990(M6.5) earthquake, Japan, *Geophys. J. Int.*, 127, 735 – 743 (1996)

Xie, M. and C.D. Lai, Reliability analysis using an additive Weibull model with bathtub-shaped failure rate function, *Reliability Eng. Sys. Safety*, **52**, 87-93 (1996)

Yen, J-Y, Chen, K-S, Chang, C-P, and Ng, S-M, Deformation and "deformation quiescence" prior to the Chi-Chi earthquake evidenced by DInSAR and groundwater records during 1995-2002 in Central Taiwan, *Earth, Planets and Space* , 58, 805-13 (2006).

Zoller G, Hainzl S, Kurths J., Observation of growing correlation length as an indicator for critical point behavior prior to large earthquakes, *J Geophys Res*, 106, 2 167–2 175 (2001).